\documentclass[aip,apl,reprint,twocolumn]{revtex4-1}
\usepackage{graphicx}
\usepackage{amssymb}
\usepackage{amsfonts}
\usepackage{amsmath}
\usepackage{amsbsy}
\usepackage{color}
\renewcommand{\vec}[1]{\mbox{\boldmath$\mathrm{#1}$}}
\let\sb=_ \catcode`\_=\active \def_#1{\ensuremath \sb{\rm#1}}

\begin{document}

\title{Time-resolved quantum spin transport through an Aharonov-Casher ring}

\author{Can Li}
%\email{lic15@lzu.edu.cn}

\author{Yaojin Li}
%\email{liyj09@lzu.edu.cn}

\author{Dongxing Yu}
%\email{yudx14@lzu.edu.cn}

\affiliation{Key Laboratory for Magnetism and Magnetic Materials of the Ministry of Education, Lanzhou University, Lanzhou 730000 , China}

\author{Chenglong Jia}
\email{cljia@lzu.edu.cn}

\affiliation{Key Laboratory for Magnetism and Magnetic Materials of the Ministry of Education, Lanzhou University, Lanzhou 730000 , China}

\affiliation{Institut f\"ur Physik, Martin-Luther Universit\"at Halle-Wittenberg, Halle (Saale) 06099, Germany}

\begin{abstract}
After obtaining an exact analytical time-varying solution for the Aharonov-Casher conducting ring embedded in a textured static/dynamic electric field, we investigate the spin-resolved quantum transport in the structure. It is shown that the interference patterns are governed by not only the Aharonov-Casher geometry phase but also the {instantaneous} phase difference of spin precession through different traveling paths. This dynamic phase is determined by the strength of applied electric field and can have substantial effects on the charge/spin conductances, especially in the weak field regime as the period of spin {precession} comparable to that of the orbital motion. Our studies suggest that a low-frequency normal electric field with moderate strength possesses more {degrees} of freedom for manipulating the spin interference of incident electrons.
%
%We investigate spin transport in a 1D Aharonov-Casher ring using algebra dynamic theory and propose an efficient method to control spin using electric field. Our work shows that the $instantaneous$ spin precession has en enormous importance on electric interference together with normal Aharonov-Casher phase. With numerical simulation, we find that perpendicular electric field can induce spin current and depolarize polarized current periodically due to the electric decoherence. Meanwhile, spin flips are seen in this system because the $instantaneous$ precession axis is various instead of being stable. Besides, the initial phase and frequency of ac electric fields can also shift the spin transportation but are inefficient than the static electric field.
\end{abstract}

\date{\today}
\maketitle
%    \makeatletter
%    \newcommand{\figcaption}{\def\@captype{figure}\caption}
%   \makeatother
%%%%%%%%%%%%%%%%%%%%%%%%%%%%%%%%%%%%%%%%%%%%%%%%%%%%%%%%%%%%%

%\section*{Introduction}

How to control and engineer the spin degree of freedom at the mesoscopic scale is a crucial step for spintronic devices \cite{Fiederling,Motsnyi,Jonker,Erve,Suzuki,Tosi,popp}. It has been demonstrated that spins of conduction electrons can be manipulated by {external gating voltage} through the Rashba spin-orbit interaction (RSOI) \cite{Mishchenko,Souma,Bercioux:2015jc,Lucignano,zhu,Nitta,Balatsky,Choi}.  Such the electric field-tunable RSOI can be achieved as well on the Aharonov-Casher (AC) effect \cite{AC} in mesoscopic ring structures \cite{Bergsten,Nitta2,Joibari}. Electron wave that traverses the AC ring along clockwise and counterclockwise directions accumulates different phases, which is reflected in the spin interference patterns of the conductance. By measuring interference patterns, the phase difference can be detected experimentally. In particular, a spin geometric phase, which is robust against the spin dephasing, can be distinguished  \cite{Nagasawa,Nagasawa2}. However, it should be noted that the spinor wave-functions used to investigate the spin interference effects in experiment and theory are \emph{not} {time-dependent} even though the spin precession in quantum transport is always there. The tilt angle between the \emph{mean} axis of the spin {precession} and the normal direction to the ring plane has been used to characterize the conductance \cite{Frustaglia:2004hd,SSQ,{Richter:2012cj}}. %based on the theoretical arguments \cite{Frustaglia:2004hd}. It should be noted that a prerequisite for a traveling electron to acquire a geometric phase is that it must move adiabatically, i.e., it should move slowly enough so that the spin stays aligned (or anti-aligned)  with the the local inhomogeneous (effective) magnetic field \cite{Diago-Cisneros}. However, true adiabaticity is very hard to achieved in experiment. The spinor wavefunctions used to investigate the spin interference effect in theory so far is not \emph{time-dependent} but characterized by the tilt angle between the mean axis of the spin precession and the normal direction to the ring plane \cite{Frustaglia:2004hd,SSQ}.

In the present study, we revisit the AC ring in the presence of static/dynamic electric fields. By giving an exact solution for traversing electrons at time $t$, a time-resolved spin precession is identified. We show that %both the charge and the spin conductance possess an AC oscillation with respect to the strength of electric fields (i.e., RSOI). However,
the interference patterns are determined by not only the AC phase but also the instantaneous phase difference of spin precessions through different traveling paths. Especially, such the time-resolved phase difference becomes more pronounced as the strength and the frequency of the applied electric field {decrease}. The spin conductivity and the bulk spin polarization (which describes the spin-dependent electronic transport in the ring) are found to be strongly depend on the spin polarization orientation of incident electrons. %, especially as the normal electric field tends to be strong and the spin precession is accelerated. The bulk spin polarization, which describes the spin-dependent electronic transport in the ring, is found to be sensitive to the spin-polarized direction of injection, and the strength and the frequency of dynamic electric fields.
Our results show that the AC ring can act as a spin interferometer, but the electric field should be properly adjusted to optimize the spin interference effects. %A slowly varying normal electric field with moderate strength is more preferable to quantum spin transport in the AC ring.

%%%%%%%%%%%%%%%%%%%%%%%%%%%%%%%%%%%%%%%%
\begin{figure}[b]
\includegraphics[scale=0.25]{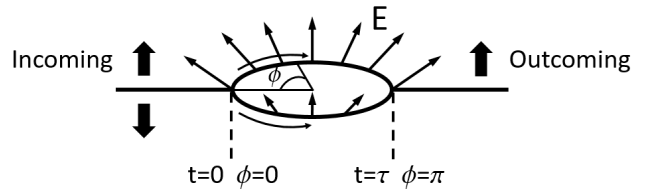}
\caption{Schematic of a quantum AC ring symmetrically coupled to two leads in the presence of electric field $\vec{E}$. A symmetrically textured electric field is assumed for the sake of theoretical investigation as that in the Ref.[\onlinecite{SSQ}].}
\label{Fig::ring}
\end{figure}
%%%%%%%%%%%%%%%%%%%%%%%%%%%%%%%%%%%%%%%%

%\section*{Time-varying wavefunction}

Let's begin with the Hamiltonian for electrons with effective mass $M$ confined to a ring of radius $a$ under a (time-dependent) textured electric field $\vec E(t) =E_r(t)\hat{\vec e}_r+E_z(t)\hat{\vec e}_z$ (cf. Fig.\ref{Fig::ring})  \cite{note},
\begin{eqnarray}
H&=&\frac1{2M}(\vec P_\phi-\frac{\mu }{2c}\vec\sigma\times\vec E)^2  \nonumber \\
&=&\frac{\vec L_{z}^2}{2Ma^2}+\frac{\mu\vec L_z }{2Mac}(\sigma_rE_z-\sigma_zE_r),
\end{eqnarray}
where we have introduced the polar angle $\phi$ in cylindrical coordinates and  $\vec L_z=-i\hbar\partial/\partial\phi $. $\sigma_i$ (with $i = r, ~ \phi, ~ z$) are the spin Pauli operators that satisfy the commute relation $[\hat \sigma_i , \hat\sigma_j] = 2i\epsilon_{ijk}\hat\sigma _k$, and $\mu=e\hbar/2Mc$ is the magnetic moment. $E_r(t)$ and $E_{z}(t)$ are assumed to be $\phi$-independent. The system then possesses the cylindrical symmetry, i.e., $[\vec{L}_z, H] =0$, which leads to the conservation of orbital angular momentum. Consequently, the invariant subspace can be labelled by certain eigenvalue $n$ of $\vec{L}_z$ and the non-autonomous Hamiltonian becomes a linear function of the $\sigma_i$,
\begin{equation}
H = \frac{\hbar\omega_0}2 n^2+\frac{\hbar\omega_r}{2}\sigma_r-\frac{\hbar\omega_z}{2}\sigma_z
\end{equation}
with $\omega_0=\frac\hbar{Ma^2}$, $\omega_z=\frac{\mu n}{Mac}E_r$, and $\omega_r=\frac{\mu n}{Mac}E_z $.
To solve the Schr\"odinger equation, $i\hbar\frac{\partial}{\partial t} |\Psi(t)\rangle= H |\Psi(t)\rangle$ without specifying the time-dependence of electric field $\vec{E}(t)$, we perform a gauge transformation \cite{WSJ,JCL},
\begin{eqnarray}
& U_g(t)=\exp[iv_z(t) \sigma_z] \exp[iv_\phi(t)\sigma_\phi], \\
& H\rightarrow {\tilde{H}}=U_g^{-1}H U_g-i\hbar U_g^{-1}\partial U_g/\partial t, \\
& |\Psi(t)\rangle\rightarrow|\tilde{\Psi}(t)\rangle=U_g^{-1}|\Psi(t)\rangle.
\end{eqnarray}
Under the best gauge conditions
\begin{equation}
\begin{split}
%
%\label{Eq::Conditions-1}
2\frac{dv_\phi}{dt}+& \omega_r\sin2v_z=0, \\
\omega_r\cos2v_\phi\cos2v_z+\omega_z &\sin2v_\phi+2\frac{dv_z}{dt}\sin2v_\phi=0,
\end{split}
\label{Eq::Conditions}
\end{equation}
we have then the diagonalized gauge Hamiltonian in the $\tilde{\sigma}_z$ representation,
\begin{equation}
{\tilde{H}}=\frac{\hbar\omega_0}{2}n^2-\frac\hbar2\frac{\omega_r\cos2v_z}{\sin2v_\phi}\tilde{\sigma}_z.
\end{equation}
Let $|m\rangle$ be the eigenstate of $\tilde{\sigma}_z$ with eigenvalue $m (=\pm1)$, the solution of the gauged Schr\"odinger equation can be written explicitly as
\begin{equation}
|\tilde{\Psi}_{n,m}(\phi,t)\rangle=e^{-i\Theta_{n,m}(t)}e^{in\phi}|m\rangle
\end{equation}
with $\Theta_{n,m}(t)=\frac1{\hbar}\int_0^t\tilde{E}_{n,m}(t')dt'$ and $\tilde{E}_{n,m}(t)=\frac{\hbar\omega_0}2n^2-\frac{m\hbar}2\frac{\omega_r\cos2v_z}{\sin2v_\phi}$ being the energy eigenvalue of the gauge Hamiltonian $\tilde{H}$.
%$\tilde{E}_{n,m}(t)=\frac{\hbar\omega_0}2n^2-\frac{m\hbar}2\frac{\omega_r\cos2v_z}{\sin2v_\phi}$.
Based on the gauge transformation $|{\Psi}(t)\rangle=U_g|\tilde{\Psi}(t)\rangle$, the real time spin-resolved solution of the original Schr\"odinger equation reads then,
\begin{equation}
|\Psi_{n,m}(\phi,t)\rangle=e^{-i\Theta_{n,m}(t)}e^{in\phi}\sum_{m'}D^{1/2}_{mm'}(v_\phi,v_z)e^{im'v_z}|m'\rangle,
\label{Eq::Psi}
\end{equation}
where
\begin{equation}
D^{1/2}[v_\phi (t),v_z(t)]
=\left[
\begin{matrix}
\cos v_\phi(t) & \sin v_\phi(t) e^{2iv_z(t)} \\
-\sin v_\phi(t) e^{-2iv_z(t)} &  \cos v_\phi(t) \\
\end{matrix}
\right]
\end{equation}
is the Wigner function.
The energy of the system is given by,
\begin{equation}
E_{n,m}(t)=\frac{\hbar\omega_0}{2}n^2-\frac{m\hbar}{2}(\omega_r\sin2v_\phi\cos2v_z-\omega_z\cos2v_\phi).
\label{Eq::Energy}
\end{equation}
It's easy to check that $|\Psi_{n,m}(\phi,t)\rangle$ are complete and orthogonal in the whole Hilbert space. The general wave-function of the ring can thus be expanded as,
$ |\Psi(\phi, t) \rangle = \sum_{n,m} C_{n,m} |\Psi_{n,m}(\phi,t) \rangle$, where $C_{n,m}$ are time-independent coefficients and completely determined by the initial conditions. It is worthy to note that $|\Psi_{n,m}(\phi,t)\rangle$ is quite general for the AC ring with any cylindrical symmetric electric field $\vec{E}(t)$. In particular, $|\Psi_{n,m}(\phi,t)\rangle$ can describe precisely and advantageously the spin {precession} in a static electric field (see below). From the best gauge conditions, Eqs.(\ref{Eq::Conditions}) with the initial values $v_\phi(0)$ and $v_z(0)$, the time-varying $v_\phi(t)$ and $v_z(t)$ can be worked out, and then all the properties of the system should be obtained.

%%%%%%%%%%%%%%%%%%%%%%%%%%%%%%%%%%%%%%%%
\begin{figure}
\includegraphics[width=8.5cm,height=11cm]{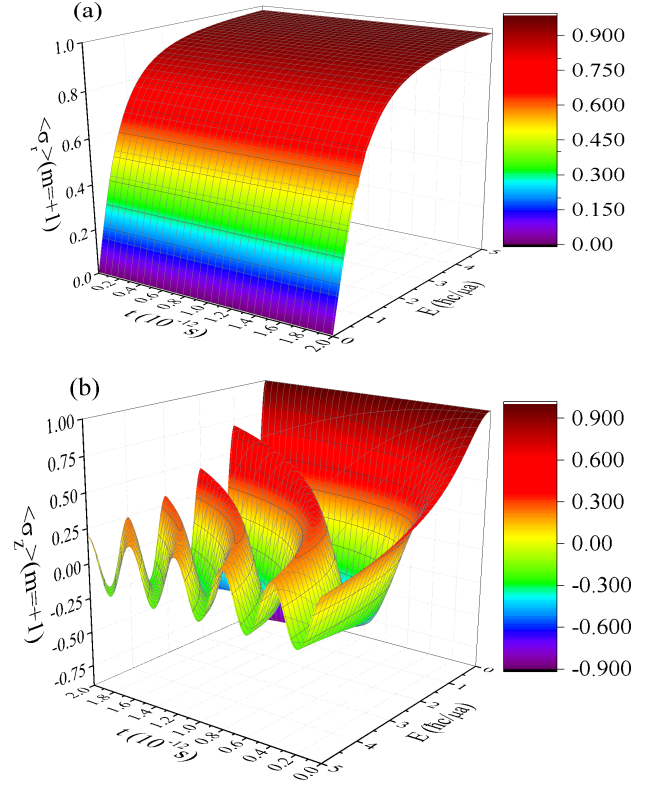}
\caption{Dynamical evolution of the spin state: (a) $\langle \sigma_r \rangle$ and (b)  $\langle \sigma_z \rangle$, respectively, in the AC ring  subjected to a \emph{static} electric field  $\vec{E} = E \hat{\vec e}_z$. The initial conditions are $v_\phi(0) = -1/2\arctan(\mu Ea/\hbar c)$ and $v_z(0) = 0$.}
\label{Fig::spin}
\end{figure}
%%%%%%%%%%%%%%%%%%%%%%%%%%%%%%%%%%%%%%%%

To get a clear insight into the physical meanings of $v_\phi(t)$ and $v_z(t)$, let's write down the expected value of spin vector $\langle \vec{\sigma} \rangle$ by using the basis $|\Psi_{n,m}(\phi,t) \rangle$,
\begin{eqnarray}
&&\langle\sigma_z\rangle = m\cos2v_\phi(t), \\
&& \langle \sigma_r \rangle= -m\sin2v_\phi(t)\cos2v_z(t), \\
&& \langle \sigma_\phi \rangle= m\sin2v_\phi(t)\sin2v_z(t),
\end{eqnarray}
which indicate that $v_\phi(t)$ describes the \emph{instantaneous} tilt angle % of spin precession
from the normal $z$-direction at time $t$ and $v_z (t)$ characterizes the spin rotation angle around the $z$-axis. In Fig.\ref{Fig::spin}, we plot $ \langle\sigma_z\rangle$ and $ \langle\sigma_r \rangle$ versus the magnitude $E$ of a \emph{static} normal electric field $\vec{E} = E \hat{\vec e}_z$ (Here it should be noted that the energy $E_{n,m}$ is time-independent even though $v_\phi(t)$ and $v_z(t)$ are time-varying under the static electric field). As one can see that $ \langle\sigma_r \rangle$ is time-independent, satisfying the conservation equation $[\sigma_r, H] =0$. When the strength $E$ of normal electric fields is enhanced, $ \langle\sigma_z \rangle$ (and $ \langle\sigma_\phi \rangle$) becomes to oscillate {precessionally} with the time $t$, % even in the presence of a time-independent electric field,
which is quite different from the previous (theoretical) spinor wave-functions that deduce a time-independent expected value of $\langle \vec{\sigma} \rangle$ once the electric field $\vec{E}$ is given \cite{Frustaglia:2004hd,SSQ}. %Clearly,  $|\Psi_{n,m} (\phi,t )\rangle$ in our case can describe advantageously the spin state at an $instantaneous$ precession time $t$.
One can also notice in Fig.\ref{Fig::spin} that,  as the normal electric field is enhanced, the spin {precession} becomes faster and the precession axis becomes to follow the direction of the effective magnetic field (along the radial direction).

%\section*{Quantum spin transport}

Based on the wavefunction $|\Psi_{n,m} (\phi,t )\rangle$, now we consider a ring symmetrically coupled to two equivalent contact leads (cf. Fig.\ref{Fig::ring}).  In clear comparison to the spin interference patterns given by time-independent spinor wave-functions with the mean axis of the spin {precession}  \cite{Frustaglia:2004hd,SSQ},  a perfect coupling between leads and ring is assumed (i.e., fully transparent contacts and no backscattering effects) to the first order linear approximation. Given an incident electron with energy $E_{F}$ and spin $|s\rangle = C_\uparrow |\uparrow \rangle + C_\downarrow |\downarrow \rangle$ ($\sum_m C_m^2 =1$)  from the left lead, depending on the spin alignment ($m$) and the direction of angular momentum (counterclockwise or clockwise with $\lambda = \pm 1$, respectively), the initial electronic state in the ring at $\phi=0$ and $t=0$ becomes a superposition of the four wavefunctions
$|\Psi_{n_m^\lambda,m} (0,0) \rangle = \sum_{m'}C_{m}D_{mm'}^{1/2}[v_{\phi}(0),v_z(0)]|m'\rangle$,
where $n_m^\lambda$ is determined by solving  $E_F = E_{n_m^\lambda,m}(0)$ in Eq.(\ref{Eq::Energy})  and do not require to be integer. Then, the incoming spin $|s\rangle$ entering the ring at $\phi=0$ propagate precessionally along the four Feynman paths and interfere at $\phi=\pi$ after time $\tau$.
To this end, we calculate the quantum probability of transmission for the outgoing spin $|s'\rangle$ channel,
\begin{equation}
T_{s'} = |\sum_{n_m^\lambda,m}\langle s'|\Psi_{n_m^\lambda,m}(\pi,\tau)\rangle|^2.
\label{Eq::T}
\end{equation}
%
%with
%\begin{equation}
%|\Psi_{n_m^\lambda,m}(\pi,\tau)\rangle = \sum_{m'}C_{m}e^{-i\Theta_{n_m^\lambda,m'}(\tau)}e^{i\lambda n_m^\lambda \pi}e^{im'v_z(\tau)} D^{1/2}_{mm'}[v_{\phi}(\tau),v_z(\tau)]|m'\rangle. \end{equation}
%
The zero-temperature charge and spin conductances are given respectively by the Landauer formula \cite{Landauer},
\begin{equation}
G_c=\frac{e^2}{h}(T_\uparrow+T_\downarrow)~ \text{and} ~ G_s=\frac{e^2}h(T_\uparrow-T_\downarrow).
\label{Eq::G}
\end{equation}
The corresponding bulk spin polarization is defined by $\ P_z={G_s}/{G_c}$ \cite{Schmidt-05}. By carrying out the tedious but straightforward algebra, we find the modulation of conductances origins indeed from the phase difference (acquired by different Feynman paths), which is however a composite of two terms: (i) the spin geometric phase accumulated by the change of spinor orientation during transport that is determined by the mean spin precession angle, and (ii) the instantaneous spin {precession} phase difference $\Delta v_z$ at $\phi=\pi$ and $t=\tau$ through different traveling paths.

%%%%%%%%%%%%%%%%%%%%%%%%%%%%%%%%%%%%%%%%
\begin{figure}[ht]
\includegraphics[scale=0.25]{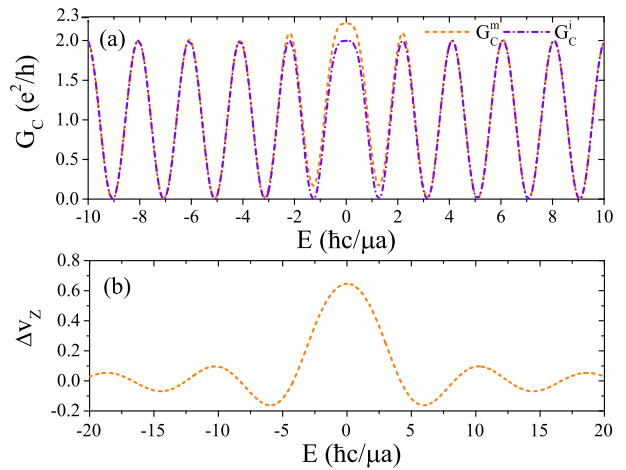}
\caption{The charge conductance with presupposed \emph{time-independent} tilt angle $v_\phi$ (dash-dot line) or self-consistently solved $v_\phi$ (dash line) as a function of the strength $E$ of normal electric field. The dash-dot curve shows good consistent with the results in Ref. \onlinecite{Frustaglia:2004hd} {except the area near E=0}. The instantaneous spin {precession} phase differences $\Delta v_z$ at $\phi=\pi$ and $t=\tau$ are shown in below as the function of $E$. Here the incoming energy is $E_F=5$ eV.}
\label{Fig::Gc}
\end{figure}
%%%%%%%%%%%%%%%%%%%%%%%%%%%%%%%%%%%%%%%%

%%%%%%%%%%%%%%%%%%%%%%%%%%%%%%%%%%%%%%%%
\begin{figure}[ht]
\centering
\includegraphics[scale=0.25]{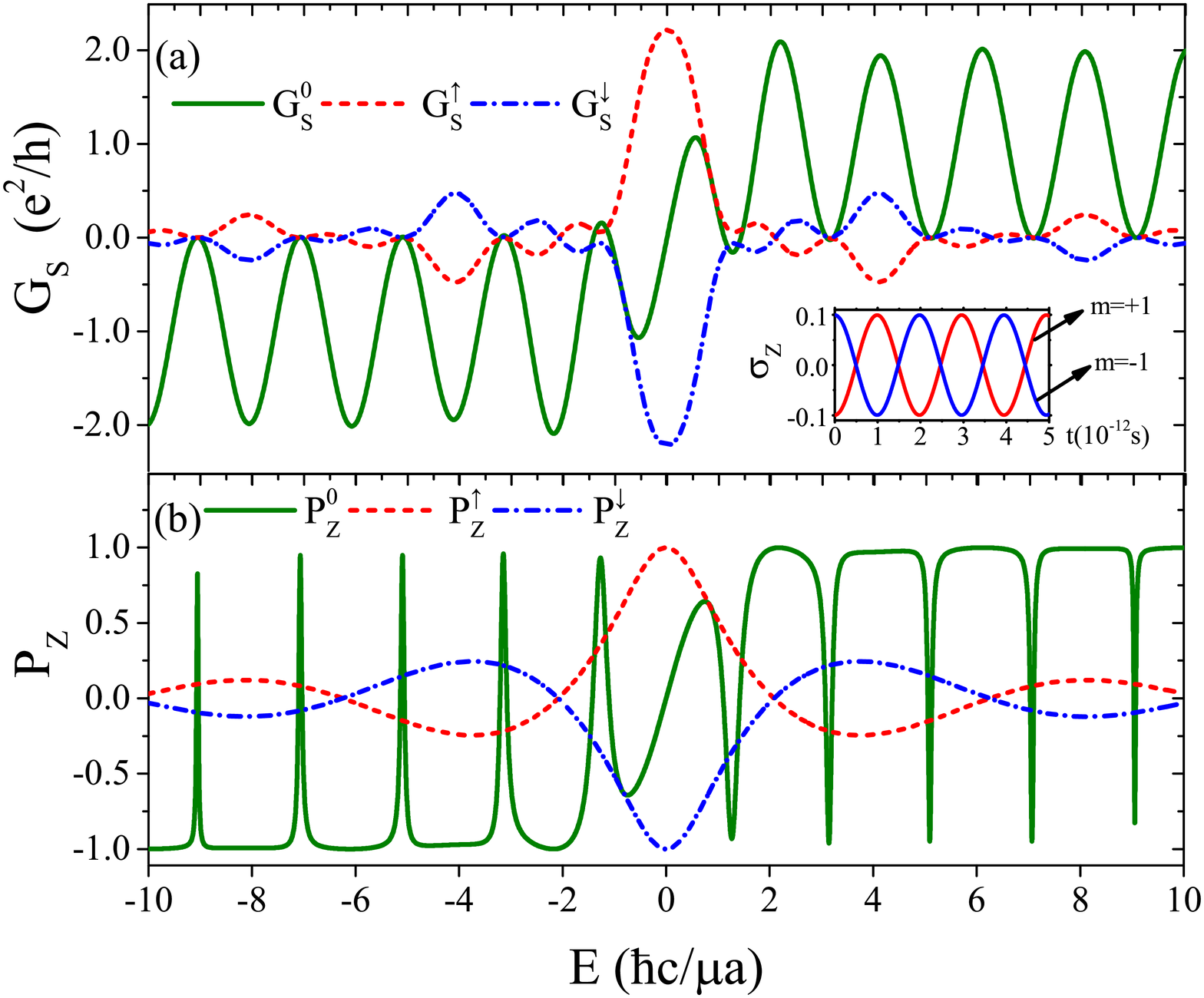}
\caption{Numerical results for the spin conductance and the spin polarization by using the time-varying wavefunction Eq.(\ref{Eq::Psi}).   $G_s^0$ ($P_z^0$), $G_s^\uparrow$ ($P_z^\uparrow$), and $G_s^\downarrow$ ($P_z^\downarrow$) correspond to the different incident spin state, $C_\uparrow=C_\downarrow=\sqrt{2}/2$, $C_\uparrow=1$ but $C_\downarrow=0$, and $C_\uparrow=0$ but $C_\downarrow=1$, respectively. The inset shows the time evolution of $\langle \sigma_z\rangle$ ($m=\pm1$) with $E=10$ $\hbar c/\mu a$. Here, $E_F=5$ eV, $v_\phi(0) = -1/2\arctan(\frac{\mu Ea}{\hbar c})$, and $v_z(0) = 0$. }
\label{Fig::G-t}
\end{figure}
%%%%%%%%%%%%%%%%%%%%%%%%%%%%%%%%%%%%%%%%

Firstly, let's try to reproduce the charge conductance, for instance that in Ref.\onlinecite{Frustaglia:2004hd}, based on Eqs.(\ref{Eq::T}) and (\ref{Eq::G}) by using the wave-function $|\Psi_{n,m} (\phi,t )\rangle$, but with a presupposed \emph{time-independent} tilt angle $v_\phi(0)=-\frac12\arctan{\frac{\mu Ea}{\hbar c}}$ \cite{Frustaglia:2004hd,SSQ}. The numerical results are shown in Fig.\ref{Fig::Gc} by the dash-dot line, which is in good agreement with the analytical expression, $G_c=\frac{e^2}{h}[1-\cos(\pi \sqrt{1+Q_R^2})]$ with $\tan2v_\phi = Q_R$. Here $Q_R$ representing the Rashba SOI constant that has the same effect as the normal electric field $E_z$ in the AC ring. However, after consistently solving the coupled differential equations Eqs.(\ref{Eq::Conditions}) with the initial conditions $v_\phi(0)=-\frac12\arctan{\frac{\mu Ea}{\hbar c}}$ and $v_z(0) =0$ (at  $\phi=0$), we find that, as the electric field $E_z$ decreases, (i) the mean axis of the spin precession tends to align itself in the normal direction of the ring plane and thus the associated spin solid angle becomes smaller; (ii) whereas, the instantaneous spin-resolved phase difference $\Delta v_z(\tau)$ at $\phi=\pi$ becomes more pronounced in weak electric field area, as the spin has a comparable {precession} period to its orbital motion and the spin-dephasing induced by the (fast) spin precession is strongly suppressed (cf. the Fig.\ref{Fig::Gc} (b)). As a result, the dynamical modulation effect of spin precession gets enhanced and the charge conductance possesses substantial deviation from the values only with the AC geometry phase (cf. Fig.\ref{Fig::Gc} (a)). On the other hand, in the presence of strong electric fields, the spin precession is accelerated (cf. Fig.\ref{Fig::spin}) and the instantaneous phase difference $\Delta v_z(\tau)$ becomes small and even random. The interference effect of the charge conductance is dominated again by the AC phase.
Such time-resolved spin precession effect also demonstrates itself in the spin transport behavior. As shown in Figs.\ref{Fig::G-t} (a), depending on the spin state of the incident electron, the spin-dependent transmission changes dramatically: one gets a large spin resistance for fully \emph{spin-polarized} incoming electron (i.e., nearly zero $G_s^{\uparrow}$ and $G_s^{\downarrow}$ in most areas of electric fields except $|E_z| < 1$ $\hbar c/\mu a$), but the similar AC oscillations of the spin conductance of the spin \emph{un-polarized} incident electron. Furthermore, Figs.\ref{Fig::G-t}{(a)} clearly indicates that the spin-polarization of the incoming electron in the AC ring can be tuned by the normal electric field (cf. also Figs.\ref{Fig::G-t} (b)). Unfortunately, the weakest RSOI realized experimentally in Ref.\onlinecite{Nagasawa} was $Q_R=0.25$, corresponding to $E=0.35 ~\hbar c/\mu a$ in our case, which is slightly higher than the point that the deviations become noticeable in Figs.\ref{Fig::Gc}. We expect a {weaker} RSOI to emphasize the spin precession effect in experiments.

%%%%%%%%%%%%%%%%%%%%%%%%%%%%%%%%%%%%%%%%
\begin{figure}[b]
\includegraphics[scale=0.27]{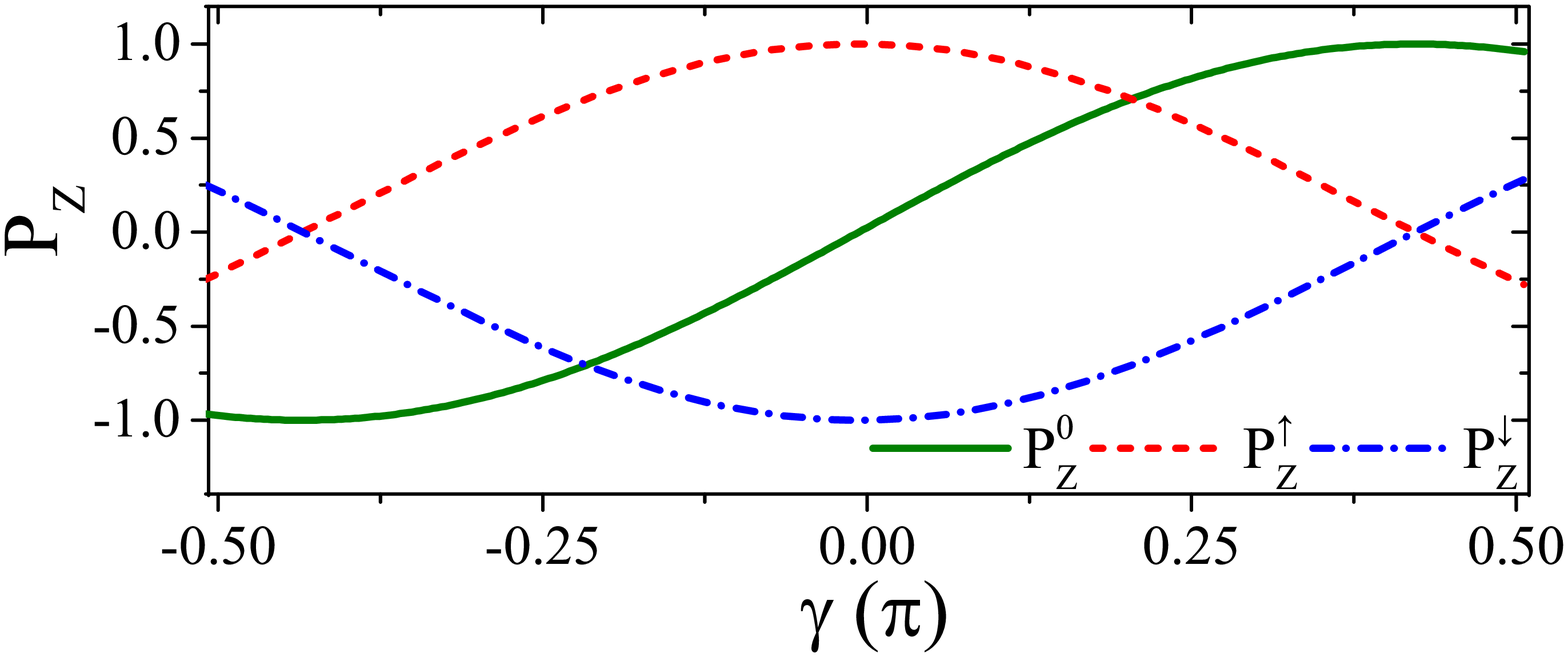}
\caption{The influence of in-plane component of the textured electric field $\vec{E} =E (\cos \gamma \hat{\vec e}_r+\sin \gamma \hat{\vec e}_z)$ on the spin polarization. The amplitude $E=4$ $\hbar c/\mu a$ is fixed during the numerical calculation.}
\label{Fig::gamma}
\end{figure}
%%%%%%%%%%%%%%%%%%%%%%%%%%%%%%%%%%%%%%%%

%%%%%%%%%%%%%%%%%%%%%%%%%%%%%%%%%%%%%%%%
\begin{figure}[t]
\includegraphics[scale=0.27]{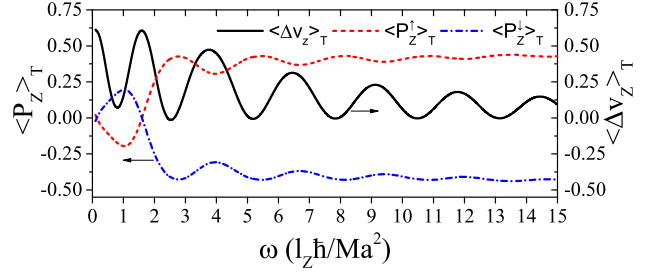}
\caption{Dynamic effects of the ac electric field  $\vec{E}(t) =E \cos(\omega t+\Phi) \hat{\vec e}_z$ on the integral-averaged spin polarization. The amplitude of applied electric field is $E=5$ $\hbar c/\mu a$ and the incoming energy reads $E_F = 5$ eV. }
\label{Fig::E-t}
\end{figure}
%%%%%%%%%%%%%%%%%%%%%%%%%%%%%%%%%%%%%%%%

The effect of an in-plane electric field is studied by tuning the tilt angle $\gamma$ of a textured electric field $\vec{E} =E (\cos \gamma \hat{\vec e}_r+\sin \gamma \hat{\vec e}_z)$. It is clear that one can not change the spin polarization orientation of incident $z$-polarized electron by the AC ring in the presence of in-plane field $E_r$ only, being consistent with the constraint condition $[\sigma_z, H] =0$ under $\gamma =0$. However, as shown in Fig.\ref{Fig::gamma}, the angle $\gamma$ is capable of controlling the modulation of polarization of electron transmitted.

To complete the discussion of dynamic effects we investigate in the following an ac normal electric field $\vec{E}(t) =E \cos(\omega t+\Phi) \hat{\vec e}_z$. Considering that not a single electron but an electron current is injected into the ring, the \emph{effective} initial phase $\Phi$  seen by each individual incident electrons at the left incoming contact changes continuously with time,  which would result in a periodic modulation in the outgoing transmission with respected to the time (equivalently, the phase $\Phi$). Therefore, we take the time integral of charge/spin conductances over a period interval $ 2\pi/\omega$. It is found that the integral-averaged spin polarization $\langle P_z \rangle$ of unpolarized incoming electron current is zero. However, the numerical results reveal that the ac field is helpful to improve the spin interference effect of the fully spin-polarized incoming electrons (cf. Fig.\ref{Fig::E-t}). Due to the dynamic phase difference $\Delta v_z$, the spin polarizations oscillates with the frequency of applied ac fields and tends to be stabilized in the high frequency region.

%\section*{Conclusion}
In conclusion, for the quantum spin transport through an AC ring in the presence of cylindrical electric fields, we have presented an exact time-dependent solution for the problem by using the algebra dynamic method and focus on the time-resolved spin interference effect. It is revealed that, besides the spin geometry phase, the instantaneous phase different of spin precession in different Feynman paths has big influence on the interference patterns in the case of weak and/or low-frequency electric fields. We have also demonstrated the possibility to control the spin polarization by the frequency, the strength, and the tilt angle of applied electric field. Our time-dependent solutions are general and can be applied to the AC ring based spintronic devices with any type of rotationally textured electric fields. In the non-ballistic regime, especially in the presence of significant disorder, the momentum scattering reorients the direction of the spin precession axis resulting in a random effective electric field and dynamic phase difference, which would lead to an average spin dephasing.

%\section*{{Acknowledgment}}
{
This work is supported by the National Natural Science Foundation of China (No. 11474138), the German Research Foundation (No. SFB 762), the Program for Changjiang Scholars and Innovative Research Team in University (No. IRT-16R35), and the Fundamental Research Funds for the Central Universities.
}


\begin{thebibliography}{99}
\bibitem{note} The correction term that contains $\sigma_{\phi}$ is neglected in the case of large angular momentum.
\end{thebibliography}
\end{document}